\documentclass[lettersize,journal]{IEEEtran}
\usepackage{amsmath,amsfonts}
\usepackage{algorithm}
\usepackage{algpseudocode}
\usepackage{array}
\usepackage{textcomp}
\usepackage{stfloats}
\usepackage{url}
\usepackage{verbatim}
\usepackage{graphicx}
\usepackage{cite}
\usepackage{hyperref}
\usepackage{xcolor}
\usepackage{harpoon}
\usepackage{mathtools}
\usepackage{multirow}
\usepackage{enumitem}
\usepackage{float}
\usepackage{colortbl}
\usepackage{diagbox}
\usepackage{booktabs}
\usepackage{threeparttable}  % 导言区添加
\usepackage[caption=false,font=footnotesize]{subfig}

\begin{document}
\title{Goal-Oriented Multi-Agent Semantic Networking: Unifying Intents, Semantics, and Intelligence}

\author{Shutong~Chen, Qi~Liao, Adnan~Aijaz, and Yansha~Deng
\thanks{S. Chen and Y. Deng are with the Department of Engineering, King’s College London, London WC2R 2LS, U.K. (e-mail: shutong.chen@kcl.ac.uk; yansha.deng@kcl.ac.uk) (Corresponding author: Yansha Deng).

Q. Liao is with the Network Automation Department, NSSR Lab, Nokia Bell Labs, 70435 Stuttgart, Germany. (e-mail: qi.liao@nokia-bell-labs.com).

A. Aijaz is with the Bristol Research and Innovation Laboratory, Toshiba Europe Ltd., Avon, Bristol, BS1 4ND,  U.K. (e-mail: adnan.aijaz@toshiba.eu).
}
}


\maketitle

%TODO 没说GSC啊
\begin{abstract}
6G services are evolving toward goal-oriented and AI-native communication, which are expected to deliver transformative societal benefits across various industries and promote energy sustainability.
Yet today’s networking architectures, built on complete decoupling of the applications and the network, cannot expose or exploit high-level goals, limiting their ability to adapt intelligently to service needs.
%However, existing networking architectures fail to support these goals due to the complete decoupling of the application and the network, which prevents the network from adapting its policies to high-level application and leads the application to make decisions without regard to real-time network conditions.
This work introduces Goal-Oriented Multi-Agent Semantic Networking (GoAgentNet), a new architecture that elevates communication from data exchange to goal fulfilment. 
GoAgentNet enables applications and the network to collaborate by abstracting their functions into multiple collaborative agents, and jointly orchestrates multi-agent sensing, networking, computation, and control through semantic computation and cross-layer semantic networking, allowing the entire architecture to pursue unified application goals.
%semantic and intent-aware mechanisms that guide coordination across layers.
%To bridge this gap, we propose a Goal-oriented Semantic Networking (GoAgentNet) architecture that enables holistic application–network collaboration by abstracting their functions into collaborative agents, and incorporates goal-oriented design to optimise the high-level goal.
We first outline the limitations of legacy network designs in supporting 6G services, based on which we highlight key enablers of our GoAgentNet design.
%, including open-domain intent interpretation, on-demand cross-layer coordination,  plug-n-play functional orchestration, and goal-oriented  multi-agent computation and communication co-design. 
%We then present the structure of the proposed GoAgentNet architecture 
Then, through three representative 6G usage scenarios, we demonstrate how GoAgentNet can unlock more efficient and intelligent services.
We further identify unique challenges faced by GoAgentNet deployment and corresponding potential solutions. 
A case study on robotic fault detection and recovery shows that our GoAgentNet architecture improves energy efficiency by up to 99\% and increases the task success rate by up to 72\%, compared with the existing networking architectures without GoAgentNet, which underscores its potential to support scalable and sustainable 6G systems.
%Finally, we demonstrate the societal benefits and sustainability of our proposed GoAgentNet through a robotic fault detection and recovery case study, where our solution improves energy efficiency by at least one order of magnitude and increases the task success rate by up to 72\%.
\end{abstract}


\section{Introduction}
% 先写IMT 2030 6G use case
% cope到UN goal
% 但是challenge是
% 传统的因为分层无法解决，下层对上层不可见
% goal-orinetned，因为是面向服务的
% 和GoC的其它magazine比
% 他们是点对点
% During the 2023 Radiocommunication Assembly, the International Telecommunication Union (ITU) has officially released the IMT-2030 framework \cite{} for 6G, which specifies six usage scenarios for future networks, including \textit{immersive communication}, \textit{massive communication}, and \textit{hyper reliable low latency communication} that extend from 5G services, and the newly introduced \textit{ubiquitous connectivity}, \textit{AI and communication}, and \textit{integrated sensing and communication}.
% These developmentsare expected to generate substantial societal impact by accelerating digital transformation across v ertical industries, and also contribute to the United Nations' Sustainable Development Goals \cite{}. 
%During the 2023 Radiocommunication Assembly, t
The International Telecommunication Union (ITU) has released the IMT-2030 framework \cite{IMT2030}, defining six usage scenarios and fifteen enabling capabilities for 6G.
These developments are expected to drive substantial societal impact across vertical industries, and support the United Nations' Sustainable Development Goals (SDG) \cite{SDG}. 
For example, immersive Extended Reality (XR) and Mixed  Reality (MR) can enhance the quality of remote education and telemedicine services (SDGs 3{\&}4: Good Health and Well Being, Quality Education); while 6G-enabled robotics can improve production efficiency (SDGs 8{\&}9: Economic Growth,  Industry Innovation and Infrastructure). 
% whereas Internet of Things (IoT) and Digital Twin enabled by ubiquitous connectivity can improve production efficiency and predictive maintenance while supporting Decent Work and Economic Growth (SDG 8) and Industry Innovation and Infrastructure (SDG 9).

To achieve these societal and sustainability goals in 6G,  the existing legacy networking architectures (e.g., OSI seven-layer model \cite{OSI}, 3GPP protocol stack \cite{3GPP}, and Intent-Driven Network (IDN) \cite{intent})  face major challenges in meeting the societal and sustainability objectives of 6G. First, traditional communication systems that transmit all source-encoded bits are approaching the Shannon limit, which is neither scalable for ultra-dense networks nor sustainable to support data-hungry applications such as XR and metaverse. Second, Service-based Architecture (SBA) using network slicing, originally designed for eMBB, URLLC, and mMTC, can hardly fulfil the diverse and coupled requirements of those new end users like robotics and XR. User intent or quality of experience (QoE) spans multiple service attributes. For example, XR requires a high data rate (570 Mbps – 4.1 Gbps) and a low interaction latency (5 – 10 ms) \cite{whitepaper}, exceeding the expressiveness of a single slice. Finally, the OSI layered design supports a hierarchical mechanism across seven layers for easy management and control. However, this isolated layer design does not allow the user intent at application layer to propagate to lower layers for cross-layer optimisation to flexibly support application goals.

To address these limitations, Goal-Oriented Semantic Communication (GSC) has emerged as a promising paradigm \cite{GoSC}. By extracting and transmitting only the semantic representation that contributes to the application goal, GSC aligns communication with task requirements while improving scalability and sustainability. 
Despite its growing interest, existing studies on GSC are mainly designed for tasks in single-domain \cite{10589469}, single-modality \cite{11008733}, or point-to-point links \cite{zhou2022task}, falling short of the multi-agent, cross-domain interactions envisioned for 6G.  
This motivates a paradigm shift towards Goal-Oriented Multi-Agent Networking, enabling coordinated orchestration of sensing, computation, networking, and control through semantics and context-aware interfaces.

%Also, the GSC designs are still limited by the rigid layered structure and lack scalability and flexibility. Although the application can provide partial guidance for network behaviours, the potential of more holistic and effective coordination between them  remains largely unexplored.
%remain fragmented and non scalable, since models developed for specific applications often fail to transfer to other tasks, due to the lack of unified service and network interfaces.
% 仍然局限于分层设计，网络和应用协同还能再explore

% 这限制 multi-agent，misalign 6G服务差异化，veritical的本质
% 另一方面， 现有解决方案仍然是孤立的点设计，不同的服务质量要求，不同的应用接口，通信协议，semantic representation, 无法互操作？

%由于缺乏这些孤立的研究缺乏统一的抽象概念，来明确如何在不同系统间公开、交换或协调任务目标、语义表示和操作需求。由于缺乏此类共享的抽象概念和交互契约，无法互操作或组合成通用架构。
%which underestimate the complexity of next-generation applications that involve cross-domain interactions, heterogeneous devices, and multi-agent orchestration. % 因此缺乏可拓展
% without the capability to support the distributed, cross-domain and multi-agent coordination required by next-generation applications running over heterogeneous devices.
% 这不align multi-hop 的 6G 服务运行在多智能体 over 异构设备的需求
% 1. 没考虑异构设备，多智能体，multi-hop, 不可拓展
% 2. semantic interfaces and cross-layer interactions 接口不明确，都是隐式的，缺乏系统级的协同能力和明确的接口定义，因此需要一个unified

Motivated by these gaps, in this paper,  we propose a Goal-Oriented Multi-Agent Semantic Networking (GoAgentNet) architecture that abstracts both network functions and application services as distributed, collaborative agents. GoAgentNet enables joint orchestration of computation and communication resources across diverse applications via semantic computation and cross-layer networking, to deliver only the minimal, most valuable information needed for goal realisation. Our main contributions are
\begin{itemize}
    \item We review the limitations of legacy architectures, including the OSI model, the 3GPP protocol stack, and Intent-Driven Network to support goal-driven 6G services.
    \item We introduce a novel Goal-Oriented Networking architecture designed to integrate seamlessly with existing architectures, comprising the Application, Agent, Knowledge, and Network Layers. Such design
    %(e.g., perceptual agent, communication agent, computation agent, actuator agent, and orchestration agent), 
    allows for  multi-agent computation and communication orchestration  via semantic computation  and   cross-layer semantic networking   towards the high-level goal.  We then identify the key challenges and potential solutions for its deployment. 
    \item Using a robotic fault detection and recovery (FDR) use case, we validate that our proposed GoAgentNet improves energy efficiency by up to 99\% and task success rate by up to 72\% compared to legacy architectures, paving the way for scalable and sustainable 6G networks. 
%We then propose a generic GoAgentNet architecture that enables open-domain intent interpretation, on-demand cross-layer coordination,  plug-n-play functional orchestration, and goal-oriented intent-aware transmission. Based on this, we further identify key challenges and outline potential solutions for its practical deployment.
    %\item We validate our proposed GoAgentNet architectures in a robotic fault detection and recovery (FDR) use case. Compared with the legacy networking architectures, our solution improves the energy efficiency by up to 99\%, and task success rate by up to 72\%.
\end{itemize}

% through establishing an application-network interface for bidirectional knowledge exchange,

\section{Legacy Networking Architecture}
In this section, we review how user intents are handled in current communications systems through the existing standardised networking architectures, including OSI seven-layer model, 3GPP protocol stack, and IDN.
\label{sec:legacy}

{\bf OSI Seven-Layer Model:}
As the foundation for conventional network architectures, the OSI Model \cite{OSI} organises network functions through a hierarchical structure spanning from Application to Physical Layer.
It enforces layer separation, under which each layer only provides pre-defined services to the layer above it, and invokes its own protocol stack to enable designated functions.
%As a result, networks only provide  transparent bit delivery at technical-level, 
As a result, networks act as an opaque data pipe carrying messages with no understanding of application goals, while applications make decisions without awareness of real-time network conditions. Such isolation prevents cross-layer optimisation and limits the network’s ability to support goal-oriented services.

{\bf 3GPP Protocol Stack Architecture:}
The 3GPP protocol stack \cite{3GPP}, as the practical realisation of the OSI model, introduces Service-Based Architecture (SBA) and Network Function Virtualisation (NFV) to enable more flexible function orchestration.
This allows network functions to expose their capability as accessible services, according to requirements of given intents, providing service differentiation.
Nevertheless, such architecture still adheres to layered boundaries, since interactions among application and network services are limited to service invocation without coordinating them toward a shared application goal.
For example, radio-level congestion control may optimise spectrum efficiency but conflict with the application's semantic or task priority.

{\bf Intent-Driven Network:}
%To mitigate the independence between the application and the network, the IDN \cite{intent} has emerged as a novel architecture that integrates application intent into network design.
%that focuses on the desired outcomes of user intent rather than mechanically regulating how the application and network should realise the intent.
The IDN \cite{intent} brings application intent into network management by translating simplified declarative statement (e.g., establish a high-speed connection between London and Paris) into concrete network requirements (e.g., minimum data rate 100 Mbps). These requirements are then mapped to low-level network policies and configurations (e.g., a QoS policy that reserves 100 Mbps along the London–Paris path) towards an intent-compliant state.
%, through which the network performs self-management and autonomous decision-making in a closed-loop manner to converge 
While this enables intent-aware automation, IDN remains limited to accommodate the diversity and complexity of 6G services.
This is because IDN still operates at the network-level, where intents are interpreted as network metrics like bandwidth or latency, and the resulting network policies focus on enforcing these requirements rather than reflecting true application's goals.
In 6G use cases, communication acts as a means to realise specific objectives of vertical domains, such as ensuring safe robotic control, rather than being a task itself. Therefore, network-level intent alone cannot ensure optimal application-layer performance.
% 只接受网络意图，只配置网络配置, 但是现在6G 的 use case里面，通信其实是一个载体，而上层任务是有自己的goal，这个goal可能不是通信的goal，通信只是为了更好的完成这个goal，所以只是网路层面的intent driven是不够先进的

\begin{figure}
\centering
\includegraphics[width=1\linewidth]{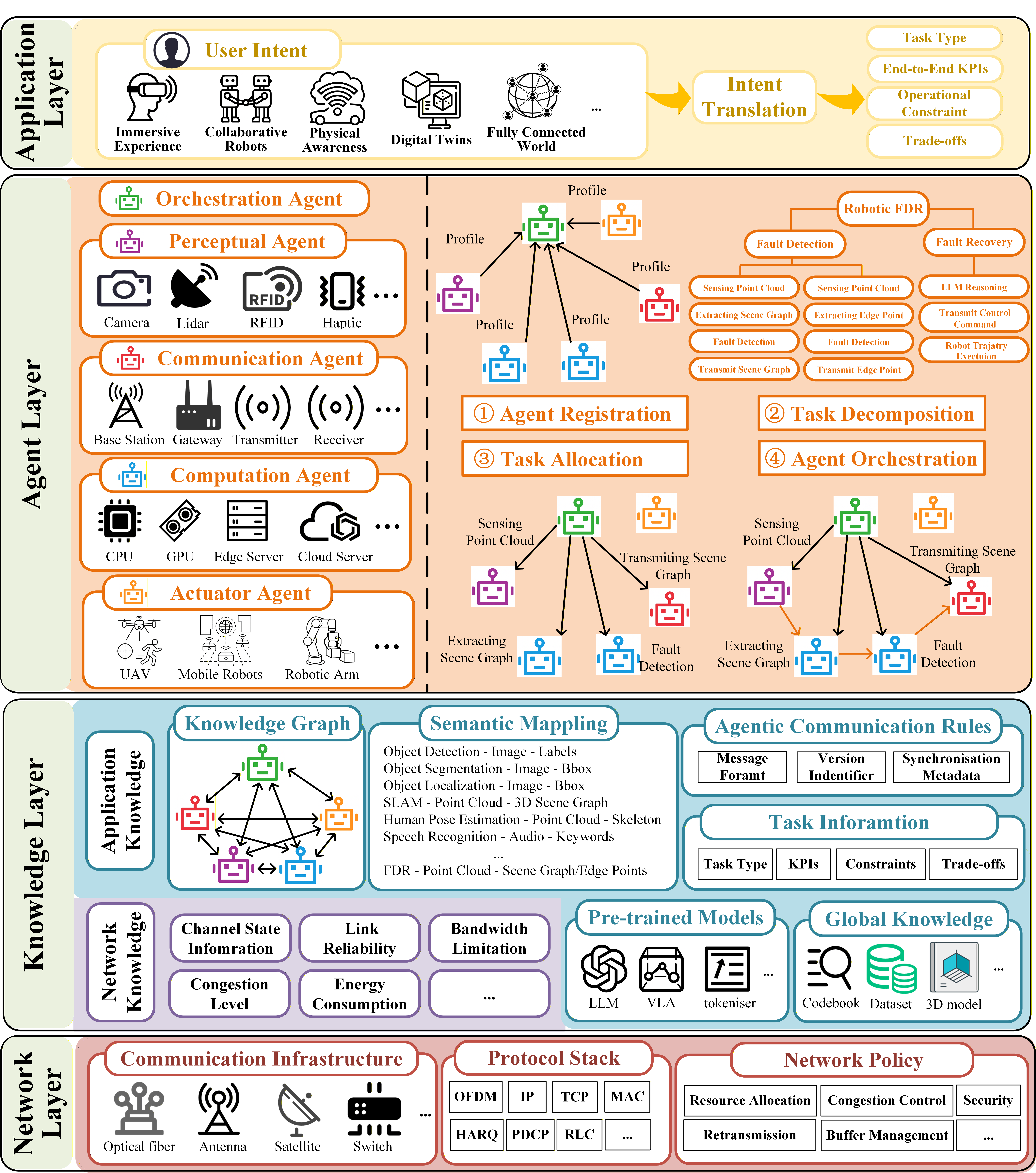}
\caption{GoAgentNet architecture and key enablers.}
\label{fig:system}
\vspace{-3ex}
\end{figure}

% knowledge layer - vertical
% agent - cross-layer conbblartion 达到最优
\section{Goal-Oriented Multi-agent Semantic Networking Architecture}
Unlike the legacy networking architecture, our GoAgentNet architecture exposes high-level application goals to lower layers and therefore aligns all layers toward an unified task objective.
%is designed to support heterogenous application across vertical domains and optimise domain-specific goals through more holistic cross-layer collaboration. 
As shown in Fig. \ref{fig:system}, GoAgentNet introduces two new virtual layers, i.e., the \textbf{Agent Layer} and the \textbf{Knowledge Layer} between the \textbf{Application Layer}  and the \textbf{Network Layer},  where the Knowledge Layer collects task information and semantic assets for cross-layer goal understanding, while the Agent Layer performs goal-oriented coordination for the application and network behaviours.
In this section, we first provide detailed explanations of the functional planes, interfaces, and cross-layer collaborations within GoAgentNet, and then present three 6G use cases to highlight how GoAgentNet effectively achieves the communication goals.

\begin{figure}[t]
\centering
\subfloat[Intent translation.\label{fig:intent}]{
    \includegraphics[width=0.98\linewidth]{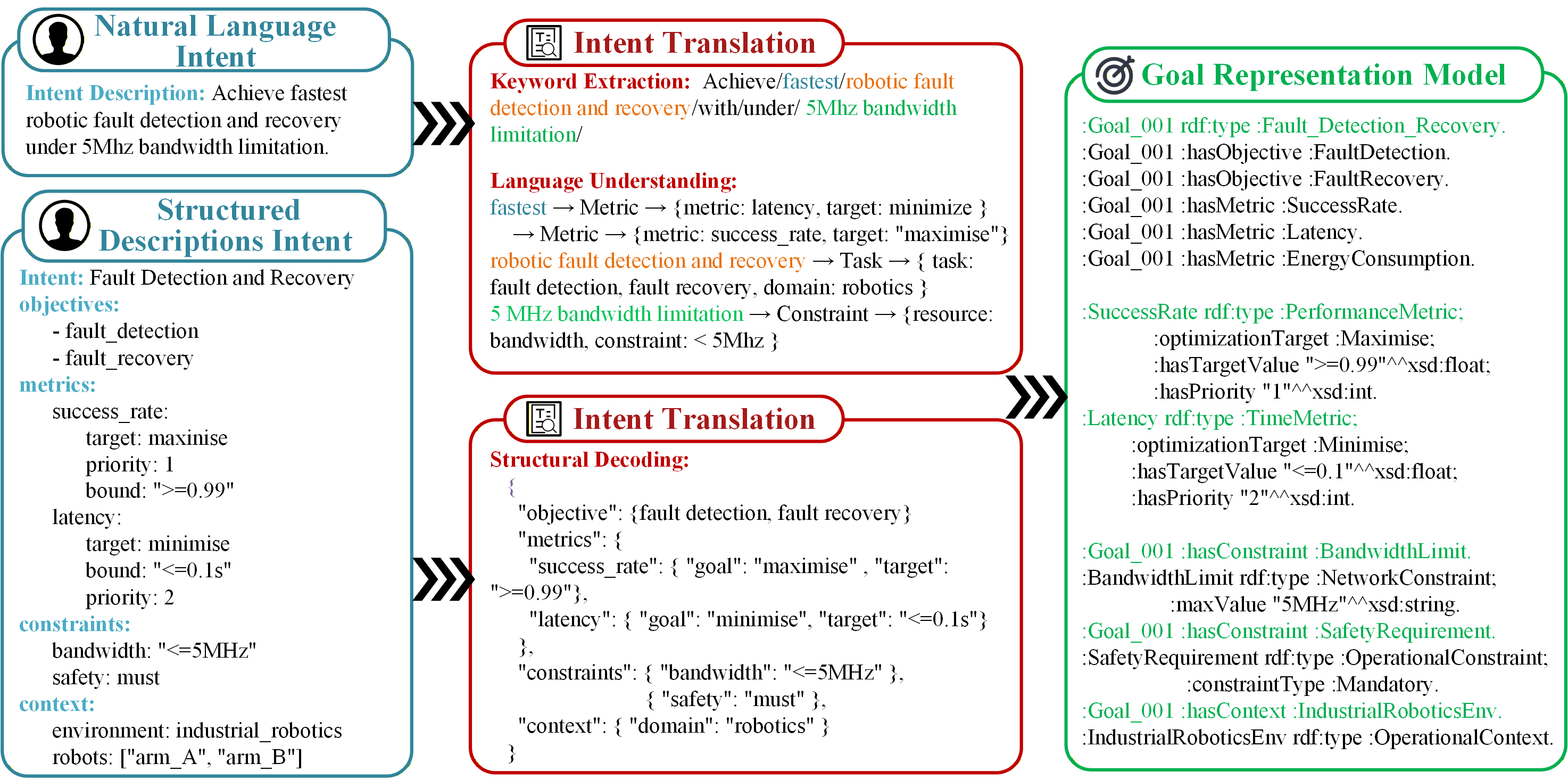}
}
\vspace{-2.2ex}
\subfloat[Knowledge graph.\label{fig:graph}]{
    \includegraphics[width=0.98\linewidth]{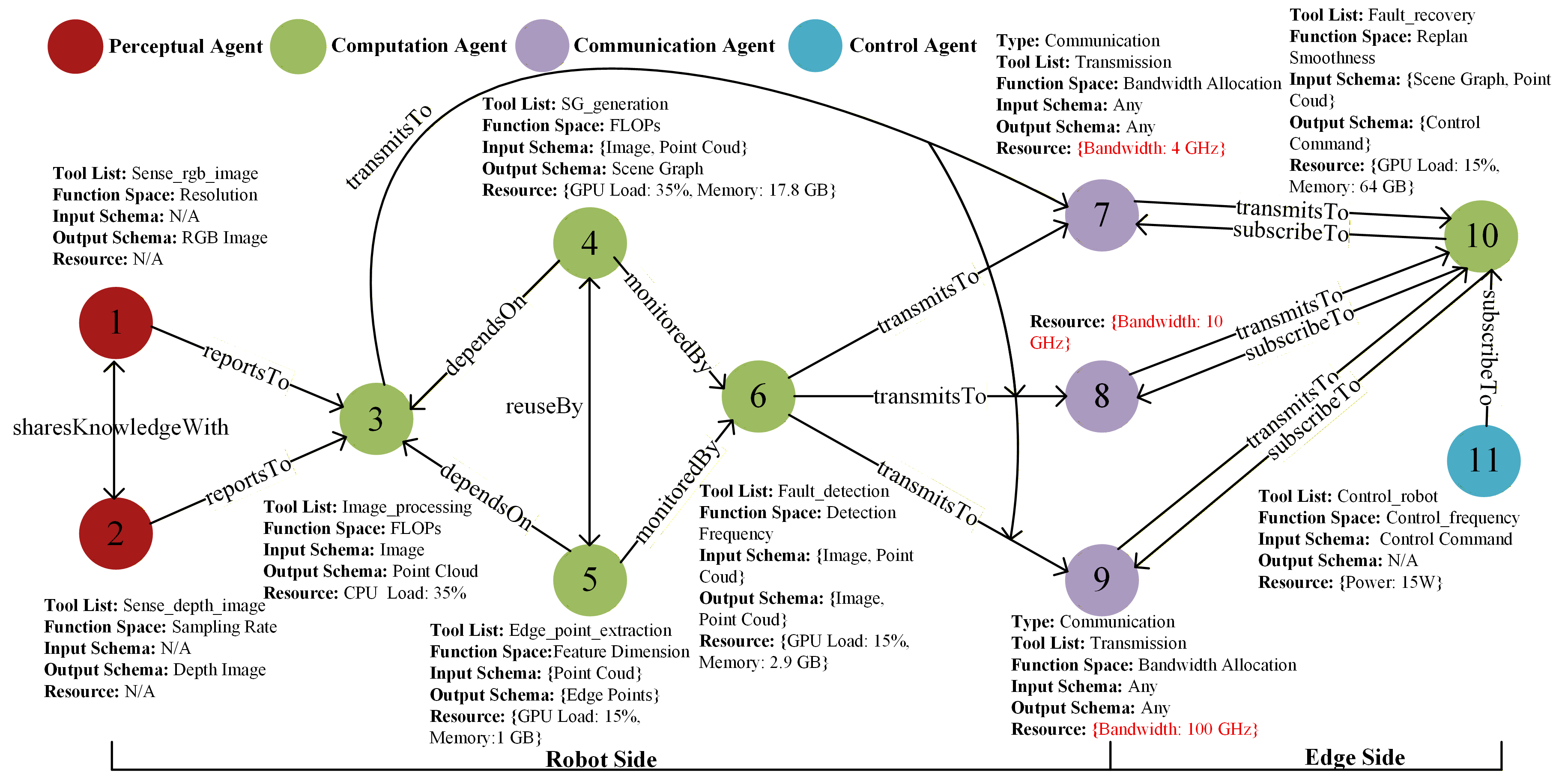}
}
\caption{Examples of intent translation and knowledge graph in our GoAgentNet architecture.}
\vspace{-3ex}
\end{figure}

\subsection{Application Layer}
As the interface between users and the network, the Application Layer is responsible for intent translation, aiming to identify the \textit{task type}, \textit{end-to-end KPIs}, \textit{operational constraints}, and  \textit{acceptable trade-offs} among KPIs.
As shown in Fig. \ref{fig:intent}, common ways to express user intent include natural language, intent-based language (e.g., Nile \cite{nile}), graphical interface, and structured domain-specific descriptions. 
When the Application Layer receives heterogeneous inputs, the \emph{intent translation module} interprets them through natural language understanding or structural decoding to extract intent-relevant labels, such as task type, KPIs, and constraints.
These labels are then encoded into a standardised goal representation model (e.g., Resource Description Framework (RDF)) for consistent goal understanding across layers, and are exposed to lower layers for cross-layer goal optimisation.
Unlike legacy networking architectures that centres on network KPIs, our intent-goal mapping goes beyond network domain, which means it also accepts intents from arbitrary vertical applications, such as robotic task success rate and positioning error.

\subsection{Agent Layer} 
GoAgentNet abstracts both application services and network functions into multiple specialised and interconnected agents, which coordinate their actions through agentic communication protocols within the Agent Layer to foster cross-layer and cross-domain collaboration.

\subsubsection{Agent Types} 
In the proposed GoAgentNet architecture, agents are divided into five types based on their functional domains, i.e., \textbf{perceptual agents}, \textbf{communication agents}, \textbf{computation agents}, \textbf{actuator agents}, and \textbf{orchestration agents},  which together cover all functions required for end-to-end orchestration of arbitrary applications, whose common function spaces are summarised in Tab. \ref{tab:function}.
\begin{itemize}
    \item \textbf{Perceptual Agents} gather sensing data through sensors, whose function space includes context-aware control over sensing behaviours.
    \item \textbf{Communication Agents} manage end-to-end data transmission and determine all network policies.
    \item \textbf{Computation Agents} perform task-specific computations (e.g., model inference and content generation), whose function space includes flexible allocation of computational resources and selection of model hyperparameters.
    \item \textbf{Actuator Agents} interact with the physical environment to execute control commands, adapting actuation dynamics as needed.
    \item \textbf{Orchestration Agents} perform dynamic task decomposition and allocation from a global perspective, reacting in real-time to unexpected disturbance, such as changing agent topology, evolving user requirements, and fluctuating network conditions.
\end{itemize}

\begin{table*}[!ht]
\caption{\centering Summary of agent types, common function entities, and corresponding function spaces.}
\label{tab:function}
\renewcommand\arraystretch{1}
\setlength{\tabcolsep}{3pt}
\begin{tabular}{|c|c|l|}
\hline
\textbf{Agent Type}                  & \textbf{Function Entity} & \multicolumn{1}{c|}{\textbf{Function Space}}                                    \\ \hline
\multirow{3}{*}{Perceptual Agent}    & Scalar Sensors           & Sampling Data Type / Sampling Rate / Measurement Range / Resolution / Averaging Window      \\ \cline{2-3} 
                                     & Vision Sensors           & Field of View /  Frame Rate / Exposure Time / Region of Interest / Colour Mode                    \\ \cline{2-3} 
                                     & Event Cameras            & Event Threshold / Polarity Mode / Timestamp Resolution / Pixel Latency / Refractory Period         \\ \hline
\multirow{4}{*}{Communication Agent} & Transport Layer          & Flow Control /  Packet Ordering / Reliability Level / Congestion Control                                                 \\ \cline{2-3} 
                                     & Network Layer            & Routing Policy /  QoS Enforcement / Gateway Selection                                                                    \\ \cline{2-3} 
                                     & Data Link Layer          & Scheduling Policy / Retransmission / Buffer Management /  Traffic Prioritisation                                               \\ \cline{2-3} 
                                     & Physical Layer           & Modulation Scheme / Coding Rate / Beamforming Configuration / Transmission Power \\ \hline
Computation Agent                    & N/A                      & Computing Precision / Floating Point Operation / CPU/GPU Frequency                                                     \\ \hline
\multirow{3}{*}{Actuator Agent}      & Robot Arms               & Joint Angle / Joint Torque / Joint Velocity / End-Effector Pose / Grasp Force                                            \\ \cline{2-3} 
                                     & Mobile Robots            & Velocity /  Heading / Turning Rate / Acceleration / E-Stop                                                                \\ \cline{2-3} 
                                     & Unmanned Aerial Vehicles & Altitude /  Thrust / Pitch / Roll / Yaw / Energy Consumption                                                               \\ \hline
\end{tabular}
\vspace{-3ex}
\end{table*}

% The Agent Layer decomposes high-level application goals into actionable subtasks, distributes them to appropriate agents according to their functional domains, enables their coordination through agentic protocols, and manages their interaction via a knowledge-guided orchestration process.

\subsubsection{Agentic Communication Protocol}
The agentic communication protocol defines how agents discover, register, and exchange information with each other and with external sources for cooperative task execution. Emerging agentic protocols include the \textit{Agent-to-Agent (A2A) Protocol}, \textit{Model Context Protocol (MCP)}, and \textit{Agent Communication Protocol (ACP)}. Although they all support inter-agent communication, we take the MCP \cite{MCP} as a representative example in the following discussion due to its open and extensible design.
 % and \textit{Agent Network Protocol (ANP)}. 

\subsubsection{Knowledge Graph Generation and Updating} 
During the system initialisation, the agent network is abstracted as a \textit{knowledge graph} stored in the underlying Knowledge Layer through agentic communication protocol. This graph is accessible to orchestration agents for coordination, and to other agents for identifying relevant peers for collaboration.
As shown in Fig. \ref{fig:graph}, nodes represent individual agents and their profiles, while the edges capture interaction links, capability dependencies, and shared knowledge.
Prior to task execution, each agent advertises its local profile that describes its tool lists (i.e., capabilities), input and output schemas, function space, and currently available resources, through standardised registration interfaces defined in MCP. This enables peers or orchestration agents to query, subscribe to, or request specific functionalities.
During task execution, agent interactions follow an MCP-defined JSON-RPC format, which specifies invoked capability, transmitted data and its type, and returned results, allowing agents to identify message content, destination, required resource, and appropriate collaborators. 
To maintain a real-time view of the agent network, the knowledge graph is dynamically updated through MCP’s built-in context synchronisation mechanisms, which broadcast state-change events when agents join, leave, or modify their capabilities and resource states.

\subsubsection{Agent Management and Orchestration}
Once the knowledge graph is established, the orchestration agents perform task decomposition and allocation across diverse agents, which enables seamless coordination between all participating entities, including application tools, communication systems, computing platforms, and control units.
High-level task is first decomposed into multiple subtasks using agent-oriented task planning methods, such as Contract Net Protocol (CNP) or Hierarchical Task Network (HTN), during which the corresponding KPIs and operational constraints are distributed to subtasks based on overall system available resources.
Each subtask is assigned to a subset of heterogeneous agents whose capabilities best match the required functions and KPI requirements. 
%while also coordinating perceptual and communication agents to decide optimal sampling rate and communication frequency under energy and sensing accuracy requirements.
The task orchestration thus transforms the high-level application goals into an execution plan represented as a traversal path or a subgraph over the knowledge graph, where the optimal collaboration path is identified using graph search (e.g., Graph-of-Thought (GoT) \cite{got}) or path optimisation algorithms (e.g., Dijkstra algorithm).
For example, the orchestration agents coordinate communication and computation agents to decide whether local processing or edge offloading achieves better energy efficiency.
This compositional approach enhances system scalability and generalization by reusing and recombination of different agents in diverse scenarios.
% The Agent Layer employs a closed-loop orchestration method, which means the task decomposition and allocation are adaptive reconfigured in response to changing network conditions or resource availability shown in knowledge graph.

\subsection{Knowledge Layer}
The Knowledge Layer is the key enabler of cross-layer collaboration in GoAgentNet. It stores both the application and network knowledge bases that all layers can utilise, which empowers the Agent Layer to formulate task-level policies, and the Network Layer to generate transmission policies.

\subsubsection{Application Knowledge Base}
By semantic-awareness reasoning over application goal and transmitted context, this base ensures that all the layers can understand, interpret, and operationalise application goals.
It comprises (a) a knowledge graph passed from the Agent Layer, (b) representation rules that regulate agentic communication, such as message format, (c) mapping rules that define the relationship between raw data, its semantic representation, and the downstream task,
%, e.g., object layout is the goal-relevant semantic representation of image data for object detection task,  
(d) pre-trained semantic models that can be accessed and reused by agents, such as codecs and tokenizers, and (e) globally shared knowledge, including codebooks, datasets, and online resources such as kinetic models for control and 3D models for generative tasks.
\subsubsection{Network Knowledge Base}
This base provides a comprehensive view of the network state to characterise the dependency between low-level network metrics and high-level application goal, thereby enabling the upper layers to react to underlying network state and make more informed computation and control decisions.
%For example, it can inform the communication agent that ``under the current channel conditions, reducing the video resolution to 720p is required to ensure transmission within 5 seconds".
It comprises feedback from the Network Layer, such as channel state information, congestion level, bandwidth limitation, link reliability, energy consumption, and security metrics.

\subsection{Network Layer}
As a native component of legacy architecture, the Network Layer retains its core role in determining network policies while being extended in GoAgentNet to align with high-level application goals. 
It transmits task-relevant semantic representations, rather than the raw  data, to accomplish a specific communication goal.
Also, it provides a cross-layer feedback loop, which reports network state to the Knowledge Layer to support the application's decision making.

\subsubsection{Goal-Oriented Semantic Communication}
Unlike legacy networks that emphasises error-free bit transmission, the Network Layer in GoAgentNet prioritizes the significance of messages relative to application goals.
It extracts the most compact, task-relevant semantic representations and transmits them only when necessary.
This makes it possible for the network to reason about what information is important, why it matters, and how to deliver it effectively.

%serves as an intelligent participant that collaborates with upper layers, rather than a passive transmission medium. 

\subsubsection{Cross-Layer Feedback Loop}
GoAgentNet establishes a bidirectional interaction loop between the Application and Network layers. 
Downward signals from the application indicate task criticality and semantic importance, allowing the network to prioritize transmissions (e.g., treating obstacle detection as high priority while background visuals as low priority in robotic pick-and-place tasks).
%In the downward direction, the application guides the network by signalling task criticality and semantic importance to enable dynamic network policies.
%For example, in a robotic pick-and-place task, when a sudden obstacle appears in the workspace, the observation related to obstacle detection must be tagged as highly critical, while updates such as background visual data can be deferred  since they are less relevant to the end-task goal.
Upward feedback reports network state to the Knowledge Layer, enabling application to predict how its behaviours (e.g., semantic encoding) influences goal realisation and adjusts accordingly (e.g., 
in the same pick-and-place example, transmitting object contours instead of full images under poor channel condition).
In extreme cases, the Network Layer can even provide explicit guidance on what data should be transmitted to maintain task continuity, such as informing the Application Layer that “under  current channel conditions, reducing image resolution to 720p is required to ensure grasping within 5 seconds”.
% it can inform the communication agent that “under the current channel conditions, reducing
% the video resolution to 720p is required to ensure transmission
% within 5 seconds”.

\subsection{Use Cases}
The proposed GoAgentNet supports diverse 6G use cases across industrial verticals. Herein we highlight three selected use cases, Robotic FDR, Vision Question Answering (VQA), and Generative AI for 3D Videos, aligned with the Physical Awareness, Collaborative Robots, and Immersive Experience families in the European Union 6G white paper \cite{whitepaper}.

\subsubsection{\textbf{Robotic Fault Detection and Recovery}} \label{sec:FDR}
FDR aims to automatically detect faults (e.g., collisions, object slippage) and generate recovery strategies for robots.
Robots transmit sensory data uplink to the edge to maintain real-time perception, where the edge server aggregates the received information and sends refined command and control data back.
This tightly coupled sensing-communication-control loop cannot be handled by legacy architectures with isolated layers.
First, such cross-domain multi-hop tasks go far beyond their capabilities since each domain is only designed to optimise its local KPIs without coordinating with each other.
Second, the lower Network Layer is unable to inform the real-time network conditions back to the upper layers, causing decisions to be based on outdated observations.
%and guide control decision-making.
%Consequently, the edge server still assumes that the knowledge collected from sensors and robots are fresh and complete when the network conditions fluctuate drastically, and makes recovery decisions based on outdated observations.

In contrast, GoAgentNet provides holistic cross-layer cooperation through agentic orchestration and closed-loop feedback empowered by the virtual Knowledge Layer.
Based on the knowledge graph, it transforms fragmented domain actions into autonomous goal-oriented orchestration, where specialised agents collaborate and negotiate to realise a unified intent while optimising their local objectives. It also facilitates closed-loop feedback that estimates how each domain’s actions influence the overall task.
In fluctuating network conditions, the edge server can promptly detect network degradation and compensate control decisions using motion prediction or channel state interpolation to preserve control stability.

\subsubsection{\textbf{Vision Question Answering}}
VQA answers user queries about an image, often requiring multiple vision tasks such as detection, segmentation, and classification. %or image reconstruction and 3D reconstruction. 
Each of these tasks has its own effectiveness-level metric, 
%for example, object detection quality depends on detection accuracy, while object segmentation is evaluated by Intersection over Union (IoU). 
yet the user intent is usually much simpler, aimed at obtaining the highest answer correctness. 
Conventional networks cannot perceive which data truly affects the answer correctness. They continue to optimise network-level KPIs, eventually stuck in a dilemma where even when these KPIs are pushed to their limits, the system still generates incorrect answers.
The GoAgentNet integrates user intent into transmission policy, selectively encoding and delivering the task-relevant semantic representations (e.g., object labels or segmentation maps) that directly contribute to answer accuracy based on its importance.
%, so that less data are transmitted while the user intent can be fully satisfied.
% 因为用户提的问题，可能是包含这些任务的，而这些任务有各种的task-level metrics，但用户的intent可能只是简单的，需要最高的回答正确率
% 传统的，什么数据才是真正影响最终答案质量，不知道怎么优化，即使把通信的KPI优化到最好，得到的answer也可能不对
% 我们的，捕捉语义，传输，不仅快，而且传输的数据是需要的

\subsubsection{\textbf{Generative AI for 3D Videos}}
Generative AI reconstructs immersive 3D videos from limited and noisy visual data through generative models (e.g., Stable Diffusion) at the receiver.
Given a user intent to achieve high-fidelity and real-time reconstruction of 3D scenes, legacy architectures suffers from significant delay and energy consumption due to bandwidth-intensive video transmission.
%First, they cannot coordinate agents toward shared reconstruction objectives due to functional separation. For example, high-fidelity reconstruction and real-time rendering are inherently conflicting from the perspective of communication agents.
%Second, they lack both reliability and flexibility to adapt to runtime changes, such as agent unavailability or resource contention.
GoAgentNet overcomes this through its goal-oriented communication design, which selectively transmits semantic representations (e.g., keypoints and text) required for reconstructing the 3D scene in generative models, while achieving better reconstruction quality, especially under severely degraded channel conditions.

\section{Challenges \& Potential Solutions}
%Despite being a promising paradigm for supporting 6G societal and sustainability  goals, GoAgentNet still faces several practical challenges in its implementation.
In this section we summarise the major challenges in implementing GoAgentNet across each layer and discuss corresponding potential solutions.

\subsection{Application Layer}
\textbf{Challenge 1: \textit{Cross-Domain Intent-Action Mapping.}} 
User intent in GoAgentNet might extend beyond communication objectives and encompass diverse vertical domains.
A fundamental challenge is how to interpret a unified goal for heterogeneous agents and map it to their own functions.
For example, in Digital Twin reconstruction, a communication agent cannot explicitly perceive how scheduling or retransmission policies affects reconstruction accuracy, since accuracy also depends on the sensing quality. When low-fidelity observations are transmitted, the communication agent cannot tell whether the performance degradation stems from transmission delay or from poor sensing by the perceptual agent.

\textbf{Solution 1: \textit{Structural Causal Model.}} 
To enable cross-domain interpretation of heterogeneous intents, the Application Layer can adopt a structural causal model (SCM) to formalise the dependencies among agent functions, domain-specific performance metrics, and global goals.
The SCM represents the cross-domain dependencies through a set of structural equations, where each equation formalises how changes in one variable influence another across domains.
For example, reconstruction accuracy can be modelled as a function of sensing quality and information timeliness, which themselves interact (higher resolution increases load; delay reduce observation fidelity).
With these structural equations, the Application Layer then performs interventional inference to identify lower bounds for domain metrics that guarantee the global objective. 
%For example, deriving information timeliness and accuracy bounds together guarantees a required reconstruction accuracy.
This enables each agent to understand how its own actions contribute to intent realisation and to adapt its behaviour accordingly.

% 每个agent function怎么影响global goal， 要达到自己domain的多少metrics，比如时延多少，才能对goal影响多少，这样所有的intent都可以被理解，因为

% 先变成high-level goal，包括metrics，contraint等。然后使用结构因果模型 (SCM) 对域之间的相互依赖性进行建模，该模型隐式地捕捉了agent的行为如何影响全局恢复效用。例如，在故障检测与恢复 (FDR) 场景中，降低通信域中的语义压缩比可能会减少传输延迟，但由于信息丢失，会降低控制智能体的重规划精度。SCM 通过结构方程形式化这种依赖关系，将域变量（感知置信度、通信保真度和控制稳定性）与系统级效用关联起来。

\subsection{Agent Layer}
\textbf{Challenge 2: \textit{Goal-oriented Coordination and Orchestration.}} 
GoAgentNet tasks are decomposed into interdependent subtasks executed by heterogeneous agents. This creates two key difficulties: (1) \emph{Capability–task mismatch}: It is unclear how to map task requirements to candidate agents, especially when multiple agents provide overlapping or partially compatible functionalities.
(2) \emph{Dynamic environments}: Agent states, resource availability, and task objectives evolve over time, while most orchestration methods assume static workflows and cannot reconfigure in real time.

\textbf{Solution 2: \textit{Utility-based Optimal Task Orchestration.}} 
%Traditional multi-agent orchestration approaches, such as game-theoretic coordination and auction-based methods, are primarily rule-driven without reasoning about the task dependencies and capability matching, thus cannot adapt to large-scale long-horizon collaborative task.
Recent efforts in Large Language Model (LLM) orchestration frameworks (e.g., GoT \cite{got}) model complex workflows as reasoning graphs that capture task dependencies and agent execution costs, enabling goal-conditioned agent selection.
GoAgentNet can build upon this by leveraging its knowledge graph, which already encodes inter-agent relationships and capability dependencies.
However, 
%unlike GoT that narrows its scope within optimising  reasoning efficiency among multiple LLMs, 
the agents in GoAgentNet are heterogeneous in capability, modality, and available resource. To support these agents, task orchestration must incorporate a \emph{unified utility function} that quantitatively evaluates not only execution cost, but also collaboration efficiency based on multiple factors, such as A2A communication cost, potential agent transfer overhead, and orchestration history.
In practice, GoAgentNet can integrate this utility-based orchestration in existing multi-agent orchestration platforms, including OpenAI’s Swarm and Microsoft's AutoGen.

\textbf{Challenge 3: \textit{Sustainable and Scalable AI for GoAgentNet.}} 
%The Agent Layer comprises numerous heterogeneous agents that collaboratively realise system-level goals. 
As the scope and diversity of high-level tasks expand, new specialised agents are introduced to support new functions. 
However, the rapid proliferation of agents would lead to expert explosion, resulting in significant model storage and maintenance overhead on computing nodes. 
More importantly, this will increase the complexity of agent coordination and orchestration, since inter-agent communication needs to be more frequent and the search space for optimal paths grows exponentially.
This highlights the need for sustainable and scalable AI in GoAgentNet, where agents must adapt, reuse, and evolve their capabilities and knowledge according to changing collaborative environments.

\textbf{Solution 3.1: \textit{Mixture of Experts.}} 
A promising solution to address the sustainability and scalability challenges is the Mixture of Experts (MoE), which decomposes a model into lightweight experts, with a gating network routing each input or task to the relevant subset. 
%The Switch Transformer developed by Google provides a practical realisation of MoE, which scales the model to trillions of experts and routes each input token to the most appropriate expert set for natural language understanding, since different experts specialise in distinct linguistic functions, such as syntax modelling, sentiment analysis, contextual reasoning, or semantic abstraction.
In GoAgentNet, MoE allows an agent to be further decomposed into finer-grained experts, which can be activated and reused on-demand across related tasks through the task-aware gating. 
For example, a speech-generation expert embedded in a video-generation agent can be invoked directly for speech-generation task without additional agent deployment.

\textbf{Solution 3.2: \textit{Weight Sharing.}} 
Weight sharing mitigates model duplication by retaining a common backbone across agents within similar domains and adding small task-specific modules.
When a new task emerges, the agent can load the pre-stored task-specific weights for rapid task switching and efficient knowledge reuse, without re-training from scratch on new tasks.
In GoAgentNet, mature implementations of weight sharing, such as Low-Rank Adaptation (LoRA) and adapter modules, can be readily leveraged to enable agents within the similar domain to self-evolve and thereby support life-long agent evolution.

\subsection{Knowledge Layer}

\textbf{Challenge 4: \textit{Multi-Modal Context and Knowledge Sharing.}} 
The Knowledge Layer serves as a shared repository that aggregates multi-modal data collected by heterogeneous agents.
However, \emph{multi-modal semantic alignment} remains challenging because information captured via different sources cannot be directly integrated into a coherent understanding of the same environment. 
%For example, sensing agents may capture the same target object through point clouds, images, and textual descriptions, which have different levels of geometric details, making it challenging to form a unified view of the object’s state and attributes.
Also, \emph{semantic ambiguity} may occur during modality transformation or knowledge exchange between agents due to distortion, incompletion, or uncertainty during data compression, transmission, or reconstruction.

\textbf{Solution 4: \textit{Unified and Shared Semantic Space.}} 
The Knowledge Layer can establish an end-to-end \emph{shared semantic space}, enabling modality-agnostic understanding.
Mechanisms such as token communication \cite{token} allow agents to transmit semantic representations or intermediate reasoning states rather than raw data, preserving fidelity and reducing ambiguity. An alternative is the KV-cache-based multi-modal communication, where agents transmit the intermediate key–value cache used for reasoning and recovering raw data rather than the raw data itself, to facilitate lossless data sharing.
In GoAgentNet, agents involved in multi-modal tasks can invoke a \emph{unified tokenizer} maintained by the Knowledge Layer to convert multi-modal data into tokens for cross-modal understanding. %and semantic alignment.
%Meanwhile, token-based processing can also mitigate semantic ambiguity because agents can leverage contextual information to infer missing or distorted data. 

% \textbf{S4.2: \textit{Semantic Representation Reuse.}} 
% For instance, a scene graph constructed for digital-twin reconstruction can provide structured labels and spatial relations that directly support object detection or tracking tasks without re-encoding raw sensory data. 

%\textbf{Solution 4.2: \textit{Semantic Fusion.}} 
%Semantic fusion integrates contextual information across modalities for robust cross-modal reasoning, thus extracting complementary semantic features that enhance modality understanding.
%For instance, to determine the robot arm grasping pose, the control agent might fuse point cloud and RGB images. When the image is partially occluded, the missing details can be complemented by depth information extracted from the point cloud, while, if the point cloud is sparse, the object contour and colour discontinuities captured by the image can refine the spatial estimation.
%Such mutual fusion compensates for modality discrepancies and promotes semantic alignment between 2D and 3D spaces by projecting geometric and visual features onto the same dimension.
%Within Knowledge Layer, semantic fusion can be deployed through cross-modal attention or graph-based aggregation mechanisms, which have been widely implemented for different modalities, such as radio signal–point cloud fusion.

% \subsection{Network Layer}
% \textbf{C5: \textit{Interoperability with Existing Infrastructure \& Standards.}} 

\section{Case Study}
\begin{figure}
\centering
\includegraphics[width=0.8\linewidth]{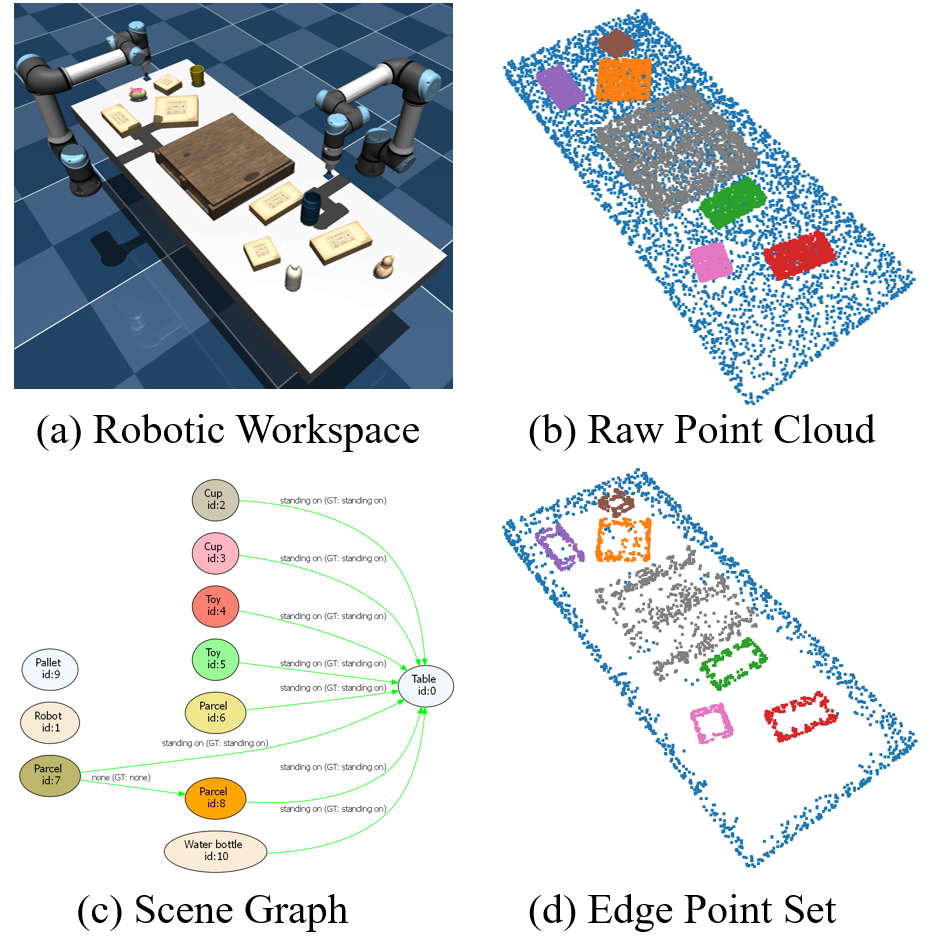}
\vspace{-2ex}
\caption{Robotic simulations and corresponding semantic representations.}
\label{fig:mujoco}
\vspace{-3ex}
\end{figure}

%To validate the effectiveness of our GoAgentNet architecture, 
We demonstrate the effectiveness of our GoAgentNet architecture in a robotic FDR task (see Sec. \ref{sec:FDR}), where a pick-and-place task is conducted within the MuJoCo simulator\cite{mujoco}.
As shown in the user intent example in Fig. \ref{fig:intent}, the user intent is first translated at the Application Layer into two FDR-oriented KPIs (i.e., task success rate and latency) and a bandwidth constraint. 
The orchestration agent then retrieves the knowledge graph  example in Fig. \ref{fig:graph} and network states at Knowledge Layer to identify an optimal execution path using \textbf{Solution 2} accordingly: 1) The perceptual agent captures the raw point cloud of robotic workspace, based on which the computation agents perform semantic extraction according to the mapping rules, supported by Transformer encoder reuse as described by \textbf{Solution 3.1}, to generate FDR suitable semantic representations, ranging from the most compact \textit{Scene Graph (SG)} Fig. \ref{fig:mujoco} (c), the lightweight object \textit{edge point set} Fig. \ref{fig:mujoco} (d), to the uncompressed \textit{raw point cloud} as Fig. \ref{fig:mujoco} (b); 2) The communication agent then selects what and when to transmit based on the bandwidth constraint, also feedbacks the network state in real-time; 3) Upon receiving these multi-modal semantic representations, a LLM computation agent reasons over them within its vocabulary book introduced in \textbf{Solution 4}, to foster unified environmental understanding and produce robot replanning strategy; 4) The actuator agent subsequently maps the replanned motions into executable control commands and applies them to the robot.
We also compare our GoAgentNet approach with a baseline that represents the legacy bit-oriented networking architecture without cross-layer collaboration and transmits the entire point cloud of the scene as Fig. \ref{fig:mujoco} (b).

%At the robot side, GoAgentNet accepts and interprets the human operator’s intent, based on which it selects one of three observations for transmission, from the most compact Scene Graph (SG) that describes workspace layouts, the lightweight object edge point set, to the complete uncompressed scene point cloud.
%At the edge side, a LLM is prompted by the received observations to generate robot trajectory replanning strategy.

% 流程，
% 然后我们的穿什么数据，baseline一直full point
% intent

\begin{figure}
\centering
\includegraphics[width=0.9\linewidth]{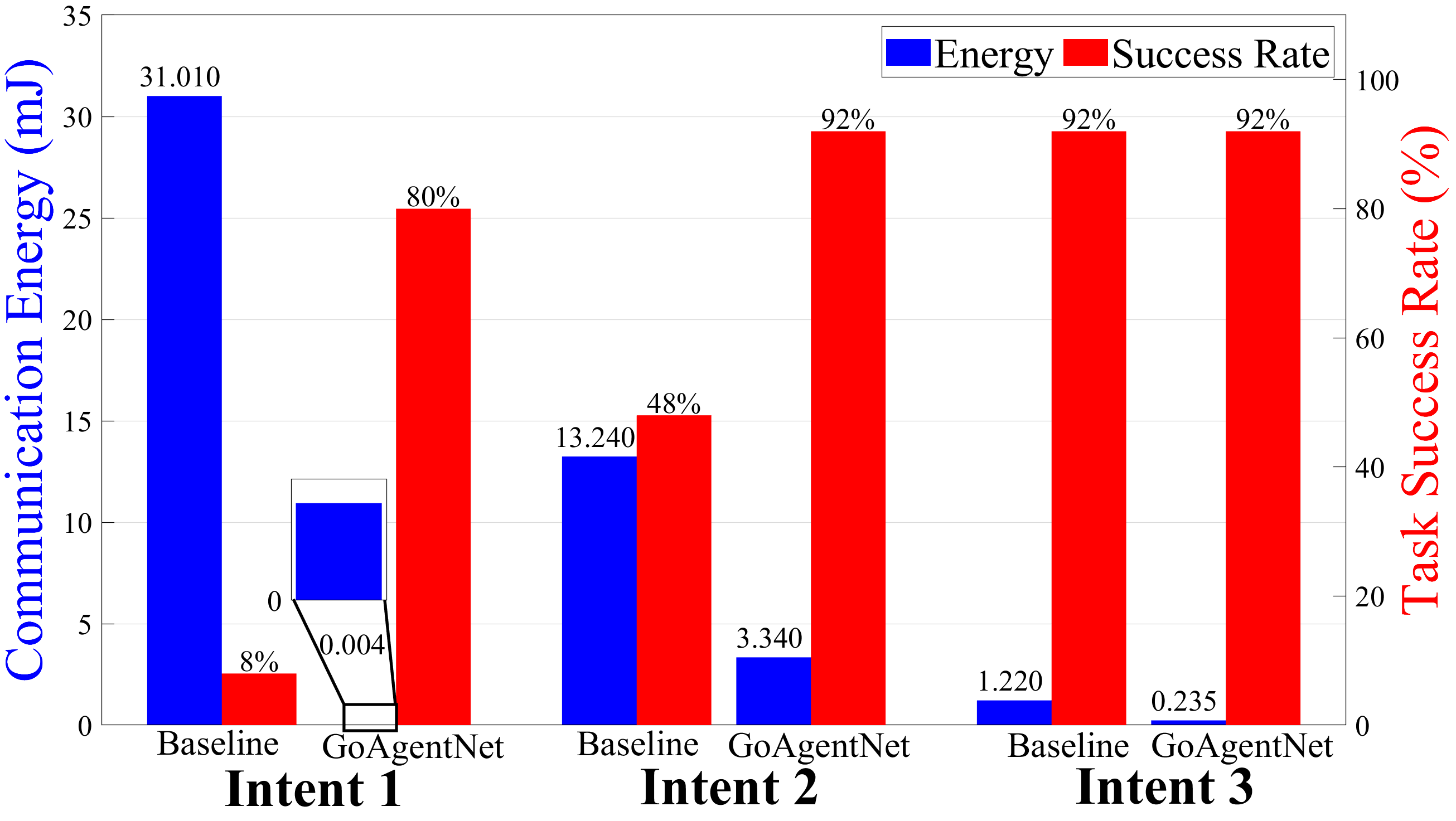}
\vspace{-2ex}
\caption{Comparison of consumed energy and task success rate between legacy and our proposed GoAgentNet networking architectures.}
\label{fig:sim}
\vspace{-3ex}
\end{figure}

Tab. \ref{tab:intent} presents how our proposed GoAgentNet adjusts both its transmitted data and orchestration paths based on three different  user intents with different bandwidth constraints, while 
Fig. \ref{fig:sim} compares the corresponding communication energy and FDR success rate between the baseline and our GoAgentNet architectures, from which we observe that:
\begin{itemize}
    \item{\textbf{Intent 1:}} When the bandwidth is below 5 MHz, our GoAgentNet  spontaneously selects the SG extraction path and prioritises the most compact yet informative SG as the semantic representation for transmission, which reduces the communication energy by over 99\% while improving the task success rate by 72\%, compared with the baseline.
    Based on its intent interpretation and closed loop feedback capabilities, it can perceive from a global view that transmitting large volume data under limited bandwidth would lead to excessive delay, ultimately compromising intent fulfilment.
    \item{\textbf{Intent 2:}} When the bandwidth constraint is slightly relaxed (below 10 MHz), GoAgentNet switches from SG to more bandwidth-intensive edge point transmission, which achieves 75\% higher energy efficiency and 44\% higher task success rate than the baseline. This demonstrates its cross-layer networking since both SG and edge points do not impose substantial delay under this bandwidth, but edge points provide richer geometric details that enable more effective FDR.
    \item{\textbf{Intent 3:}} When bandwidth becomes sufficiently large (below 100 MHz), GoAgentNet still upload edge points, which reduces the communication energy by 80\%, while maintaining the same task success rate as the baseline.
    This is because although the complete point cloud is transmittable in this setting, the edge points already contain sufficient information for task completion, making other task-irrelevant transmission unnecessary.
\end{itemize}
These observations again highlight the overall effectiveness 
of GoAgentNet as a socially beneficial and sustainable communication paradigm.
Through multi-agent orchestration and cross-layer semantic networking, it fulfils intents that extend beyond the communication domain, and also transmits minimal but informative data to promote energy sustainability.

\begin{table}[]
\centering
\caption{Intent-Aware Agent Orchestration.}
\setlength{\tabcolsep}{3pt}
\renewcommand\arraystretch{1}
\begin{tabular}{|m{4cm}|m{2.5cm}|>{\centering\arraybackslash}m{1.5cm}|}
\hline
\multicolumn{1}{|m{4cm}|}{\centering \textbf{Intent}} &
\multicolumn{1}{m{2.5cm}|}{\centering \textbf{GoAgentNet Orchestration Path} (See Fig.~\ref{fig:graph})} &
\multicolumn{1}{m{1.5cm}|}{\centering \textbf{Transmitted data}} \\
\hline
1. Achieve the highest task success rate for robotic FDR under a \textbf{5MHz} bandwidth constraint. &
1 $\rightarrow$ 2 $\rightarrow$ 3 $\rightarrow$ 4 $\rightarrow$ 6 $\rightarrow$ 7 $\rightarrow$ 10 $\rightarrow$ 11 &
Scene Graph \\
\hline
2. Achieve the highest task success rate for robotic FDR under a \textbf{10MHz} bandwidth constraint. &
1 $\rightarrow$ 2 $\rightarrow$ 3 $\rightarrow$ 5 $\rightarrow$ 6 $\rightarrow$ 8 $\rightarrow$ 10 $\rightarrow$ 11 &
Edge Points \\
\hline
3. Achieve the highest task success rate for robotic FDR under a \textbf{100MHz} bandwidth constraint. &
1 $\rightarrow$ 2 $\rightarrow$ 3 $\rightarrow$ 5 $\rightarrow$ 6 $\rightarrow$ 9 $\rightarrow$ 10 $\rightarrow$ 11 &
Edge Points \\
\hline
\end{tabular}
\vspace{-3ex}
\label{tab:intent}
\end{table}

% \begin{table}[t]
% \centering
% % \setlength{\abovecaptionskip}{-0.01cm} 
% \caption{Simulation Setup}
% \renewcommand\arraystretch{1.2}
% \begin{tabular}{|p{0.5\columnwidth}|p{0.4\columnwidth}|}
% \hline
% Intent  & Orchestration Path (See Fig. \ref{fig:graph}) \\ \hline
% Achieve the highest task success rate for robotic fault detection and recovery under a 5 MHz bandwidth constraint.   & 1               \\ \hline
% Achieve the highest task success rate for robotic fault detection and recovery under a 10 MHz bandwidth constraint.  & 2               \\ \hline
% Achieve the highest task success rate for robotic fault detection and recovery under a 100 MHz bandwidth constraint. & 3               \\ \hline
% \end{tabular}
% \end{table}

\section{Conclusion}
In this article, we proposed a generic Goal-Oriented Multi-Agent Semantic Networking (GoAgentNet) architecture, which abstracts the application services and network functions into cooperative agents, and orchestrates multi-agent sensing, communication, networking, and control towards the same high-level goal through semantic computation and cross-layer semantic networking.
We first outlined the limitations of existing network architectures to support 6G societal and sustainability goals.
Based on this, we detailed the structure and key enablers of our proposed GoAgentNet architecture.
We also identified the main challenges and potential solutions in practical deployment of GoAgentNet.
Finally, our robotic fault detection and recovery (FDR) case study showcased that our GoAgentNet architecture can significantly improve the task success rate with much lower energy consumption.
As 6G aims to contribute to the Sustainable Development Goals of the United Nations and empower vertical industries with diverse and coupled goals, the proposed GoAgentNet architecture lays a solid foundation for 6G ecosystem that facilitates environmental sustainability and delivers long-term societal benefit.

\bibliographystyle{IEEEtran}
\bibliography{IEEEabrv,Magazine_Netork}

\end{document}